\documentstyle[12pt,psfig]{article}

\begin{document} 

\title{Antikaon Production in Proton-Nucleus 
Reactions and the $K^-$ properties in nuclear matter\thanks{Supported 
by Forschungszentrum J\"ulich}}
\author{A. Sibirtsev\thanks{On leave from the Institute of 
Theoretical and Experimental Physics, 117259 Moscow, Russia.} \ 
and W. Cassing \\
Institut f\"ur Theoretische Physik, Universit\"at Giessen \\
D-35392 Giessen, Germany}
\maketitle

\begin{abstract}
We calculate the momentum-dependent potentials for $K^+$ and $K^-$ mesons
in a dispersion approach at nuclear density $\rho_0$ using the 
information  from the vacuum $K^+ N$ and $K^- N$ scattering 
amplitudes, however, leaving
out the resonance contributions for the in-medium analysis. Whereas the
$K^+$ potential is found to be repulsive ($\approx$ + 30 MeV) and to show 
only a moderate
momentum dependence, the $K^-$ selfenergy at normal nuclear matter density
turns out to be $\approx$ - 200 MeV
at zero momentum in line with kaon atomic data, however, 
decreases rapidly in magnitude for higher momenta. The antikaon production
in p + A reactions is calculated within a coupled transport approach and
compared to the data at KEK including different assumptions for the 
antikaon potentials. Furthermore, detailed predictions 
are made for $p + ^{12}C$
and $p + ^{207}Pb$ reactions at 2.5 GeV in order to determine 
the momentum dependent antikaon potential experimentally.
\end{abstract}

\vspace{2cm}
\vspace{1cm}

\section{Introduction}
One of the most exciting ideas regarding the hadron
properties in  nuclear matter is the modification of their 
mass or energy as a result of their interactions with the nuclear 
environment or a partial restoration of chiral symmetry.
According to effective chiral Lagrangians the in-medium antikaon mass
should be substantially reduced while the kaon mass is expected to
be slightly enhanced \cite{Kaplan,Waas1,Schaffner,BrownRho}. 
The experimental studies for kaonic 
atoms~\cite{Friedman} indicated that even at normal nuclear 
density the $K^-$-mass might decrease by $\simeq 200$~MeV 
which would imply    kaon condensation at 2-3 $\rho_0$.
The attractive antikaon potentials also have a strong 
impact on subthreshold $K^-$ production. Indeed, a dropping 
of the in-medium hadron mass  shifts the production threshold to lower
energies which leads to an enhanced production yield.

Antikaon enhancement in nucleus-nucleus collisions has been seen in
GSI experiments~\cite{Schroter,Barth,Senger} on strangeness production. 
The dynamical studies performed in 
Refs. \cite{Li,Cassing1,Li97,Cassing2}  indicate that  a dropping 
of the $K^-$ mass by $\simeq 100-120$~MeV (at nuclear saturation density)
is necessary to explain the $K^-$ production data for $Ni + Ni$
at 1.6 and 1.85~A$\cdot$GeV. These potentials are roughly half of those
found for kaonic atoms in \cite{Friedman}. 
Note that the experiments with kaonic atoms investigate stopped
$K^-$-mesons while in the heavy-ion collisions at GSI  
antikaons with momenta from 300 - 800~MeV/c (relative to the
baryonic fireball) are probed. Thus the measurements might be compatible 
if the antikaon potential would show up strongly momentum  dependent. 
On the other hand, in nucleus-nucleus collisions a 
complex dynamical time evolution of the system is involved and explicit
momentum-dependent potentials cannot be extracted from a comparison of
calculations to the experimental data. It is thus desirable to obtain 
information about the $K^+$ and $K^-$ potentials under circumstances where
the dynamics are much better under control and where 
the momentum dependence can be studied explicitly. This 
should be the case for proton-nucleus
reactions; here, first transport calculations 
with antikaon potentials have
been presented in Ref. \cite{Varenna}.

The in-medium $K^-$-nucleon scattering amplitude and 
$K^-$-mesons mass in nuclear matter in the presence of the
${\Lambda}(1405)$ resonance was studied in 
Refs.~\cite{Waas,Koch,Lutz}. We note that the ${\Lambda}(1405)$ 
resonance amplitude fits the $K^-N$ scattering for low antikaon momenta
relative to the baryonic matter ( $p_K \le 300$~MeV/c) as will also
be shown in the following. Furthermore, the $K^+$-nucleon scattering
amplitude has been investigated experimentally in nuclei by measurements
of the elastic and inelastic scattering of kaons on $^{6}$Li 
and $^{12}$C in Ref. \cite{Peter1}.  

In this work we investigate more systematically the momentum dependence
of the $K^+$ and $K^-$ potentials at finite nuclear density and its
relevance for strangeness production in 
proton-nucleus reactions. We start 
in Section 2 with a reminder of the elementary production 
cross sections from
$pN$ and $\pi N$ production channels and evaluate the $K^+$ and $K^-$
potentials in a dispersion approach in Section 3. Detailed 
calculations for
antikaon spectra are presented in Section 4 in 
comparison to the scarce data
taken at KEK \cite{Chiba}. Section 5, furthermore, 
contains our predictions
for antikaon sprectra at 2.5 GeV on $^{12}C$ and $^{208}Pb$ targets, 
which can be measured experimentally at
COSY, whereas Section 6 concludes this study with a summary and discussion
of open questions.  

\section{Elementary $K^-$ production cross sections}
The calculations on strangeness production from either
proton-nucleus or heavy-ion collisions are very 
sensitive to the input for the elementary processes. 
In case of $p + A$ reactions these important ingredients are
the $\pi N \to K^- X$ and $pN \to K^- X$ channels as
well as a selfconsistent description of the $K^-$-meson
propagation in the residual target nucleus due to the very strong 
$K^-$-nucleon final state interaction. 

\begin{figure}[h]
\psfig{file=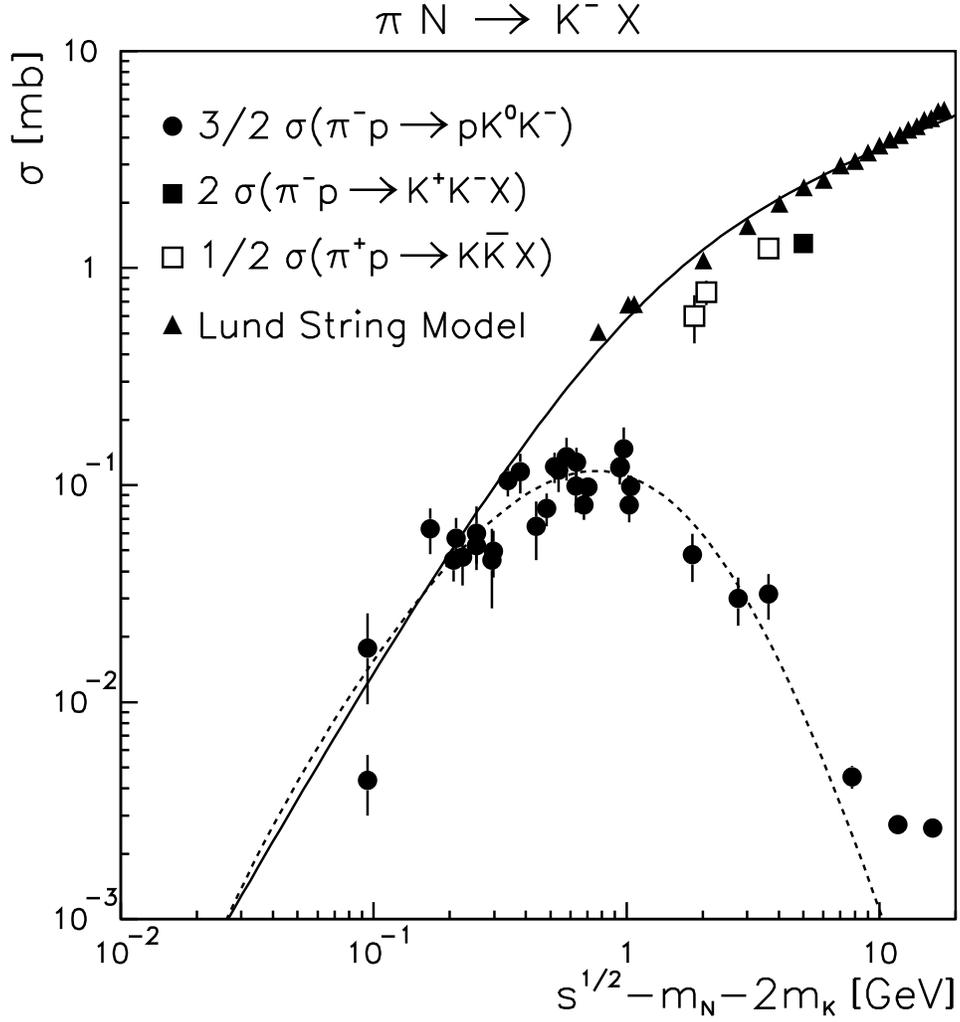,width=15cm}
\caption[]{\label{kmin7} The $K^-$-production cross section from 
$\pi N$ collisions as a function 
of the excess energy $\epsilon = \sqrt{s} - m_N - 2 m_K$ 
above the $\pi N \to N K {\bar K}$ 
reaction threshold. The dashed line shows the exclusive cross 
section  from Ref.~\protect\cite{SiKoCa} while the solid line 
is our parametrization for the inclusive $K^-$-production; the triangles 
show the calculations within the Lund-string-model (LSM) at higher energy
while the experimental data (full dots and open squares)
are taken from Ref.~\protect\cite{LB}.}
\end{figure}

Recently, the exclusive antikaon production from both
$\pi N$ and $NN$ collisions has been analyzed in terms
of the One-Boson-Exchange-Model (OBEM)~\cite{SiKoCa}; it was
found that the available experimental data for
the $\pi N \to N K {\bar K}$ reaction can be
reasonably fitted by a $K^{\ast}$ exchange model. In the latter work
the average $\pi N \to N K K^-$ cross section
has been parameterized as
\begin{equation}
\label{piex}
{\bar \sigma}(\pi  N \to N K K^-)=1.6815 \
\left(1-\frac{s_0}{s} \right)^{1.86} \
\left(\frac{s_0}{s} \right)^{2} ,
\end{equation}
where the cross section is given in $mb$, $s$ is the
squared invariant energy and $\sqrt{s_0}=m_N+2m_K$,
with $m_N$ and $m_K$ denoting the masses of the nucleon and kaon,
respectively. The relations between the different isotopic channels 
of the reaction $\pi  N \to N K {\bar K}$ are explicitely given in
Ref.~\cite{SiKoCa}.  The average cross section~(\ref{piex})
is shown in Fig.~\ref{kmin7} by the dashed line together
with the experimental data~\cite{LB} corrected 
by the isospin factors from  Ref.~\cite{SiKoCa}. 
The data are shown as a
function of the excess energy $\epsilon = \sqrt{s}-\sqrt{s_0}$
above the reaction threshold and are well reproduced 
by the parameterization~(\ref{piex}).

Note that at excess
energies above the $\pi$-meson mass  reactions channels with
additional pions become possible which have to be taken into
account for the inclusive $K^-$-meson production. At excess
energies of about 2 GeV the inclusive $K^-$ production cross section
(open squares) is about an order of magnitude higher 
than the exclusive one
(full dots). In order to match the inclusive cross 
section at high energies
we adopt the Lund-String-Model (LSM) \cite{LUND}. 
The calculated cross section
for the reaction $\pi N \to K^- X$ from the LSM is shown
by the triangles in Fig.~\ref{kmin7}.  Experimental data are   
available for the $\pi^- p \to K^+K^-X$ and 
$\pi^+p \to K {\bar K} X$ partial cross sections~\cite{LB}, which
can be related to the inclusive $K^-$-meson production
cross section via isospin symmetry as (see also Ref.~\cite{Waters}):
\begin{equation}
\sigma (\pi N \to K^- X)= 2\sigma (\pi N \to K^+ K^- X)
=\frac{1}{2} \sigma (\pi N \to K {\bar K} X). 
\end{equation}
These inclusive cross sections (open and 
full squares in Fig.~\ref{kmin7}) 
are below the LSM
calculations within a factor of 1.5 - 2. Note, 
however, that reactions with
antihyperons, i.e.  
$\pi N \to {\bar Y} K^- X$, also contribute
to the inclusive antikaon production at excess energies
$\sqrt{s}-m_N-2m_K >1.6$~GeV, which are taken into account in the LSM.

The solid line in Fig.~\ref{kmin7} illustrates our parameterization 
for the inclusive $\pi N \to K^- X$ cross section:
\begin{equation}
{\bar \sigma}(\pi  N \to K^- X)= 1.45
\left( 1- \frac{s_0}{s} \right)^2
\left( \frac{s}{s_0} \right)^{0.26}
\end{equation}
where the cross section is given in $mb$ and $s_0=3.709$~GeV$^2$.

\begin{figure}[hbt]
\psfig{file=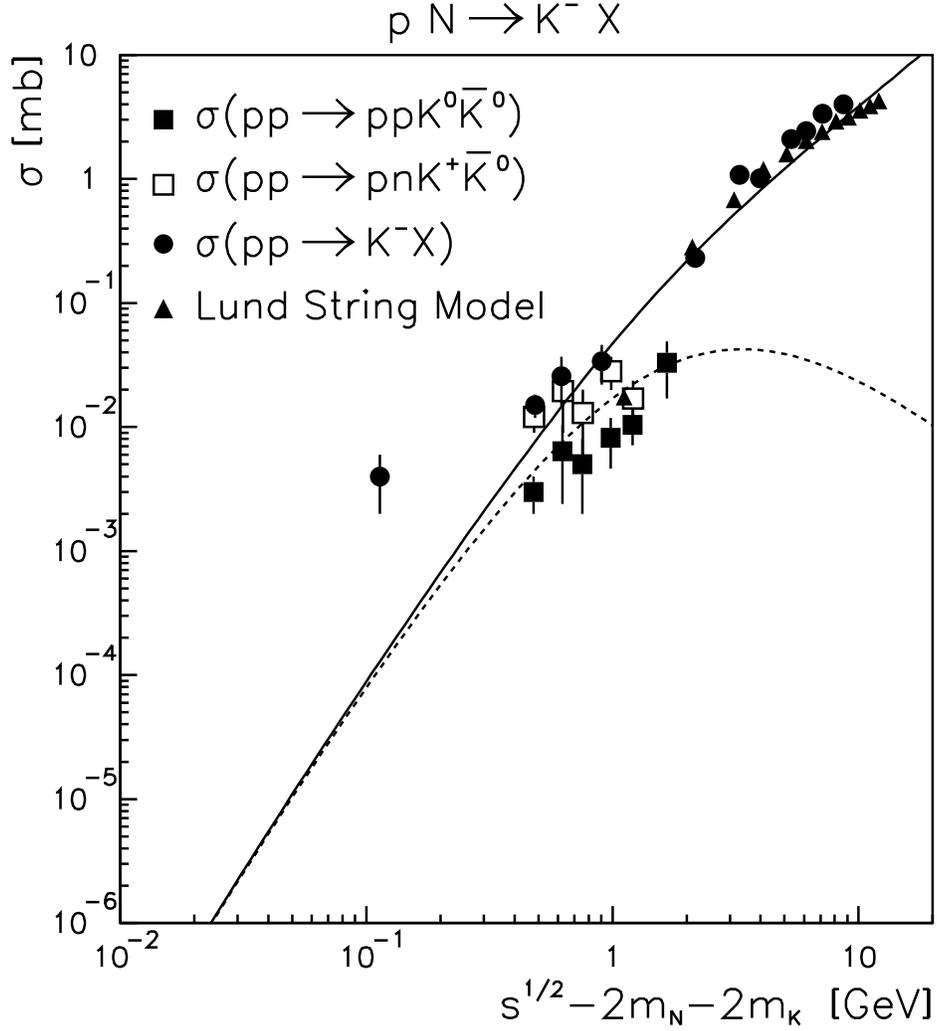,width=15cm}
\caption[]{\label{kmin8}The cross section for $K^-$-production
in $pN$ collisions as a function of the excess energy above the 
$N N \to N N K {\bar K}$ reaction threshold. The dashed line 
shows the exclusive cross section
calculated within the OBEM~\protect\cite{SiKoCa} while the solid line 
is our parametrization for the inclusive $K^-$-production cross section. 
The triangles show the calculations within the LSM in comparison to
the experimental data (full dots) taken from 
Ref.~\protect\cite{LB}.}
\end{figure}

The cross section for the $NN \to NN K {\bar K}$ reaction,
with $N$ being either a proton or neutron, $K=K^+$ or $K^0$
and ${\bar K}=K^-$ or ${\bar K^0}$, has been calculated within the
OBEM taking into account both pion and kaon 
exchange~\cite{SiKoCa}. The isospin averaged $pN \to NNKK^-$
cross section is shown by the dashed line  in 
Fig.~\ref{kmin8} together with the available experimental data
for the $pp\to pp K^0 {\bar K^0}$ and $pp\to pn K^+ {\bar K^0}$
reactions~\cite{LB}. The full dots in Fig.~\ref{kmin8} 
show  the cross sections for the inclusive $K^-$-meson production from 
$pp$ collisions~\cite{LB,NUC}. The triangles represent
the results from the LSM, which are well in line with the experimental
data at high energies. The single data point at an 
excess energy of $\simeq 0.1$~GeV
falls out of the systematics and violates severly a scaling with 4-body
phase space at low excess energies.

The solid line in Fig.~\ref{kmin8} shows our
parameterization for the inclusive $pN \to K^- X$ cross section:
\begin{equation}
{\bar \sigma}(p N \to K^- X)= 0.3
\left( 1- \frac{s_0}{s} \right)^3
\left( \frac{s}{s_0} \right)^{0.9}
\end{equation}
with the cross section given in $mb$ and $s_0=8.19$~GeV$^2$.

In order to propagate the antikaons through nuclear matter
we account for the elastic and inelastic $K^-N$
processes and implement the corresponding cross
section from the compilation of experimental data~\cite{LB}. Since
the elastic and inelastic cross sections with nucleons are rather well
known, antikaon rescattering and absorption is 
expected to be well under control.
For an explicit representation of these cross sections we refer 
the reader  to Ref. \cite{Cassing1}. 

\section{Dispersion approach for kaon and antikaon potentials}
Due to strangeness conservation the 
${\bar K}$-meson production is associated with  
$K$-produc\-tion~\footnote{Fig.~\ref{kmin7} indicates 
that at high energies also antihyperons might be produced 
instead of kaons.} and we have to account for the in-medium 
modifications of both, kaons and antikaons. Furthermore,
to study the $K^-$-momentum spectra from  $pA$ collisions 
one has to evaluate the density as well as the momentum 
dependence of the $K^{\pm}$ potentials in nuclear matter. 

As pointed out by Brown~\cite{Brown1} 
the study of kaonic atoms indicates a decrease of 
the $K^-$ mass by $200 \pm 20$~MeV~\cite{Friedman},
while the experiments at GSI suggest that it is sufficient
to drop the antikaon mass at normal 
baryon density $\rho_0$ by $\simeq$100-120~MeV~\cite{Cassing1,Li97}.
This conflicting results might be explained when
assuming that the $K^-$ potential or selfenergy depends on the antikaon 
momentum $p_K$ relative to the nuclear matter. Indeed,
the investigations with the kaonic atoms deal with 
antikaons practically at rest whereas the studies at GSI 
probe the $K^-$-mesons with
$300\le p_K \le 600$~MeV/c in the nucleus-nucleus cms. 

For a first order estimate of the kaon and antikaon 
potentials we use the low
density approximation and express the $K$ and 
${\bar K}$-potential as a function of the baryon 
density ${\rho}_B$  
and the $K^{\pm}$-momentum $p_K$ relative to 
the nuclear matter rest frame as
\begin{equation}
\label{pot1}
{\Pi}^{\pm}({\rho}_B,p_K)= -4 \pi \  {\rho}_B \  Re f^{\pm}(p_K) ,
\end{equation}
where $Re f^{\pm}$ is the real part of the $K^+N$ or $K^-N$
scattering amplitude.

The relation between the $K^+ N$ and $K^- N$ scattering amplitudes is
given by the crossing symmetry
\begin{equation}
\label{sym}
f^{\pm}(\omega)=f^{\mp \ast}(-\omega)
\end{equation}
and the forward scattering amplitude in the laboratory can be written as
\begin{equation}
\label{amp}
f^{\pm}(\omega)= D^{\pm}(\omega) + iA^{\pm}(\omega) ,
\end{equation}
with $\omega = \sqrt{p_K^2+m_K^2}$, where $p_K$ and $m_K$
are the laboratory momentum and kaon mass, respectively, while
the imaginary and real part of the amplitude are related 
by the dispersion relation.

Note that the difficulty in the calculation of the
real part for the forward scattering amplitude by
dispersion relations comes from the large unphysical cut 
in the $K^-N$ scattering amplitude and from the 
available data on the total
cross sections in the physical region.

With the substraction point at $\omega =0$ the real 
part of the $K^{\pm} N$
amplitude may be written in the form~\cite{Amati,Perrin}
\begin{equation}
\label{def}
D^{\pm} = D(0) \ \mp  \ R_Y^{\pm}(\omega) \ + \ 
\lbrace I^{\pm}(\omega < m_K) +  I^{\pm}(\omega > m_K) \rbrace .
\end{equation}
The second term in (\ref{def}) stems from the 
contribution from $\Lambda$ and
$\Sigma $ poles ~\cite{Queen},
\begin{equation}
\label{pole}
R_Y^{\pm}(\omega ) =  \frac{\omega}{16 \pi  \ m_N^2} \ 
\left( \frac{g^2_{N \Lambda K} \  [ m_K^2-(m_{\Lambda} -m_N)^2 ]} 
{{\omega}_{\Lambda} \ (\omega \pm {\omega}_{\Lambda})} + 
\frac{g^2_{N \Sigma K } \  [ m_K^2-(m_{\Sigma}-m_N )^2 ]}
{{\omega}_{\Sigma} \ (\omega \pm {\omega}_{\Sigma}) } \right) ,
\end{equation}
where $m_N$, $m_{\Lambda}$ and $m_{\Sigma}$ stand for the nucleon, 
$\Lambda$ and $\Sigma$-hyperon masses, respectively.
$g^2_{N \Lambda K }$ and $g^2_{N \Sigma K}$ are the coupling 
constants, while the  kaon energy ${\omega}_Y$ corresponds to
the process $Y \leftrightarrow K^- N$, i.e.
\begin{equation}
{\omega}_Y = \frac{m_Y^2-m_N^2-m_K^2}{2m_N} .
\end{equation}

As was proposed by Bailon et al.~\cite{Bailon} and 
Hendrick and Lautrup~\cite{Hendrick}, rather than 
to determine both coupling constants it is sufficient to
use the effective coupling constant 
\begin{equation}
g^2=\frac{1}{4 \pi} \left( g^2_{N \Lambda K} +
g^2_{N \Sigma K } \frac{ m_K^2-(m_{\Sigma}-m_N )^2}
{m_K^2-(m_{\Lambda}-m_N )^2} \right)
\end{equation}
and approximate the $\Lambda$- and $\Sigma $-poles by a single
pole located between the two, i.e. replacing Eq.~(\ref{pole}) by
\begin{equation}
R_Y^{\pm}(\omega ) = g^2  \ \frac{\omega}{{\omega}_Y \
(\omega \pm {\omega}_Y )} \ 
\frac{m_K^2-(m_{\Lambda}-m_N )^2}{4m_N^2}
\end{equation}
with ${\omega}_Y=0.11$~GeV.

In Eq. (\ref{def}) the third term represents the integral 
contribution from both the unphysical $\omega < m_K$~\cite{BRMartin} 
and the physical 
region $\omega > m_K$. The amplitude from the unphysical 
region can be obtained from the low energy solution of Kim~\cite{Kim} or 
A.D.~Martin~\cite{ADMartin} as
\begin{equation}
\label{lowen}
I^{\pm}(\omega < m_K) = \mp \frac{\omega}{\pi m_N}
\int_{{\omega}_{\Lambda \pi}}^{m_K} 
\frac{d{\omega}' \ (m_N^2+m_K^2+2m_N{\omega}' )^{1/2}}
{{\omega}' \ ({\omega}' \pm \omega )} \  A^-({\omega}')
\end{equation}
where ${\omega}_{\Lambda \pi}$ is the kaon energy corresponding
to the $K^- N \to \Lambda \pi $ reaction threshold,
\begin{equation}
{\omega}_{\Lambda \pi} = \frac{ (m_{\lambda}+m_{\pi})^2-
m_N^2-m_K^2}{2 m_N} ,
\end{equation}
and $ A^-$ denotes the imaginary part of the $K^-N$
forward scattering amplitude.

The contribution from the ${\pi}^0 \Lambda $ energy up to the threshold 
for $K^-p$ elastic scattering stems from two resonances:
the $\Sigma (1385)$, which couples to the $\pi \Lambda$ ($88 \pm 2 \%$)
and $\pi \Sigma$ ($12 \pm 2 \%$) channels, and 
the $\Lambda (1405)$ that decays entirely through the $\pi \Sigma $ 
channel. A recent review on the phenomenological
extrapolation of the $K^-N$ scattering amplitude below the
threshold is given in Refs.~\cite{Miller,Barrett}. 

In the present calculations we adopt the  effective range approximation
and express the $s$-wave amplitude as 
\begin{equation}
\label{below}
A^{-}(\omega )= 
\frac{0.5 \ b_0}{(1+a_0q)^2+b_0^2q^2} + 
\frac{0.5 \ b_1}{(1+a_1q)^2+b_1^2q^2} ,
\end{equation}
where $q$ is the imaginary kaon momentum in the $KN$ 
center-of-mass system, 
\begin{equation}
q^2= \frac{m_N^2 (m_K^2-{\omega}^2)}
{m_N^2+m_K^2+2m_N{\omega} },
\end{equation}
and the complex scattering lengths $a_j+ib_j$ are taken  from 
the solution of Kim~\cite{Kim}. The first term in 
Eq.~(\ref{below}) is the cut-off at the $\pi \Sigma $ threshold 
while the second term is that at the $\pi \Lambda $ threshold.

The contribution from the physical region is calculated as
\begin{equation}
\label{princ}
I^{\pm}(\omega > m_K) = \frac{\omega}{4 {\pi}^2}
P \int_{m_K}^{\infty} \frac{d{\omega}' \ \sqrt{{{\omega}'}^2-m_K^2}}
{{\omega}'} 
\left[ \frac{{\sigma}^{\pm}({\omega}')}{{\omega}'-\omega }
-\frac{{\sigma}^{\mp}({\omega}')}{{\omega}'+\omega } \right],
\end{equation}
where ${\sigma}^{\pm}$ stands for the total $K^+N$ and
$K^-N$ cross section. In the following we will
calculate the $K^+p$ and $K^-p$ amplitudes~\footnote{Actually
it is quite difficult to calculate the $K^-n$ and $K^+n$ amplitude
since there are no experimental data on the relevant
total cross sections at low energies~\cite{LB} as well as
the information about the  unphysical cut in $K^-n$ scattering,
which complicates the determination of the coupling constants
and the substraction~\cite{Hendrick}.}
using the cross sections $\sigma^{\pm}(\omega)$ parameterized as 
described in the Appendix.

\begin{figure}[hbt]
\psfig{file=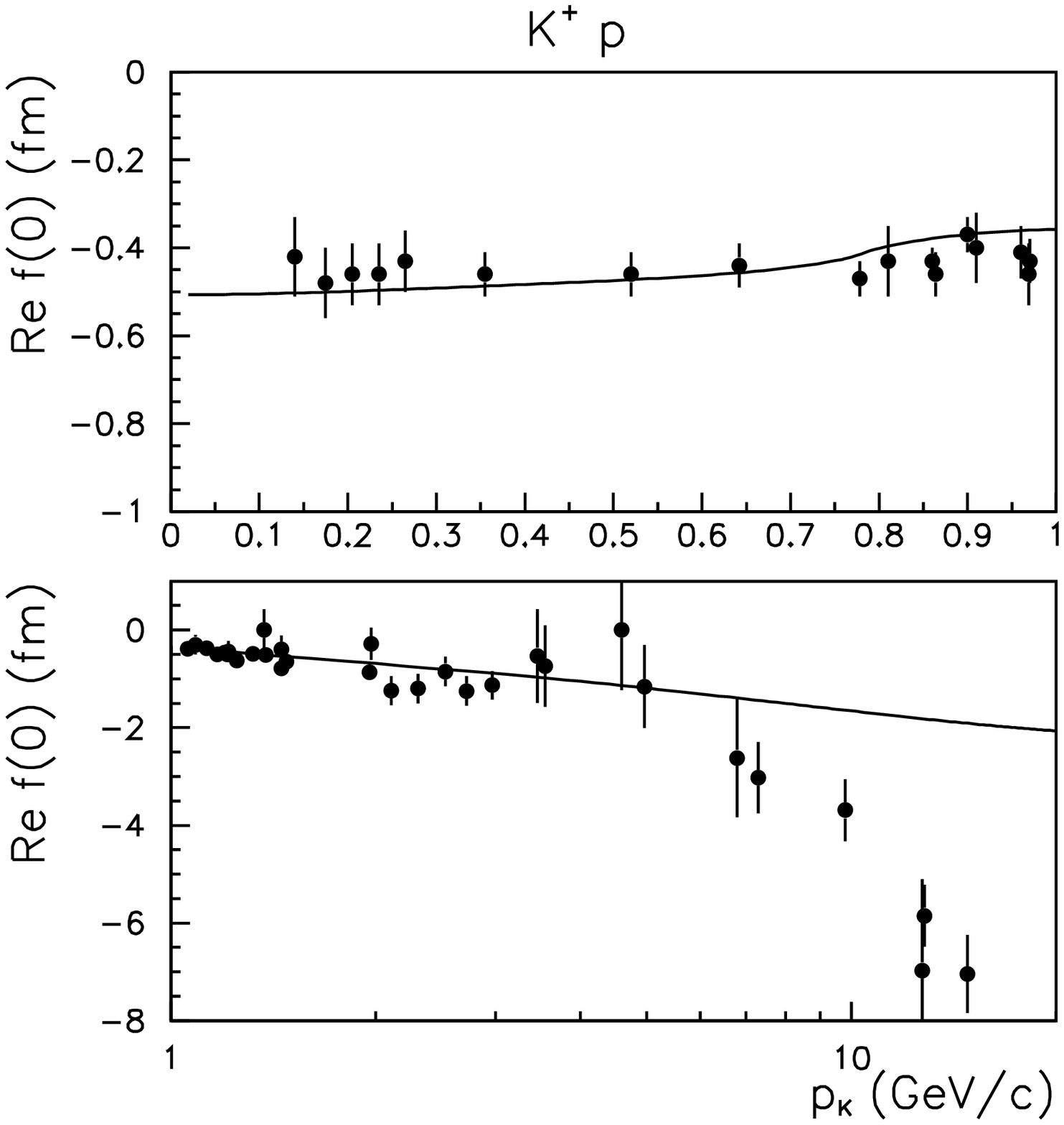,width=15cm}
\caption[]{\label{kmin4}Real part of the $K^+p$
scattering amplitude (solid line) in comparison
to the experimental data taken from Ref.~\protect\cite{Dumbrais}.}
\end{figure}

Fig.~\ref{kmin3} shows the real part of the forward $K^-p$ 
scattering amplitude calculated by Eq.~(\ref{def}) together 
with the experimental results collected 
by Dumbrais, Dumbrais and Queen~\cite{Dumbrais}. 
Note that the sign of the real part of the forward scattering
amplitude is undetermined experimentally in most cases,
but can be fixed from the analysis of the dispersion 
relations~\cite{Queen1,Dumbrais1}. We find a 
quite reasonable agreement between our dispersion calculations 
and the data up to very high kaon momenta. The real part
of the $K^+p$ amplitude is shown in Fig.~\ref{kmin4} correspondingly.
Note that dispersion calculations systematically disagree 
with the data for kaon momenta above 5~GeV/c. 

At low energy the $K^- p$ interaction in vacuum is repulsive due to the
coupling to the $\Lambda (1405)$ resonance. Hovewer, the observations
of kaonic atoms~\cite{Friedman} show that the real part of the $K^-$
optical potential in nuclear matter is attractive. Indeed, if the
$\Lambda (1405)$ is a quasibound state of the ${\bar K}N$ system, then
it should dissolve in the medium due to the Pauli
blocking of the proton in the $\Lambda (1405)$.
A recent analysis of the low energy ${\bar K}N$ interaction
in baryonic matter has been performed by Waas, Rho and Weise~\cite{Waas}
and the  in-medium properties of the $\Lambda (1405)$ resonance 
were investigated by Ohnishi, Nara and Koch~\cite{Koch}.
Both investigations show an attractive ${\bar K}N$
interaction at nuclear matter density. Furthermore, the
mean-field approach of Brown and Rho~\cite{Brown} accordingly
assumes that in nuclear matter the $\Lambda (1405)$ resonance
is insignificant and has no relevance at nuclear densities.
We thus also neglect the low energy solution~(\ref{lowen}) 
and simply drop the contribution from both $\Sigma (1385)$
and $\Lambda (1405)$ resonances.

It is convenient to write the modification of the $K^{\pm}$-meson 
energy in the medium in terms of a potential $U^{\pm}$ via
\begin{equation}  
\label{forder}
U^{\pm}= - \frac{2 \pi}{m_K} \  {\rho}_B \ Re f^{\pm} ,
\end{equation}
where ${\rho}_B$ is the baryon density and $m_K$ the bare kaon mass. 

\begin{figure}[hbt]
\psfig{file=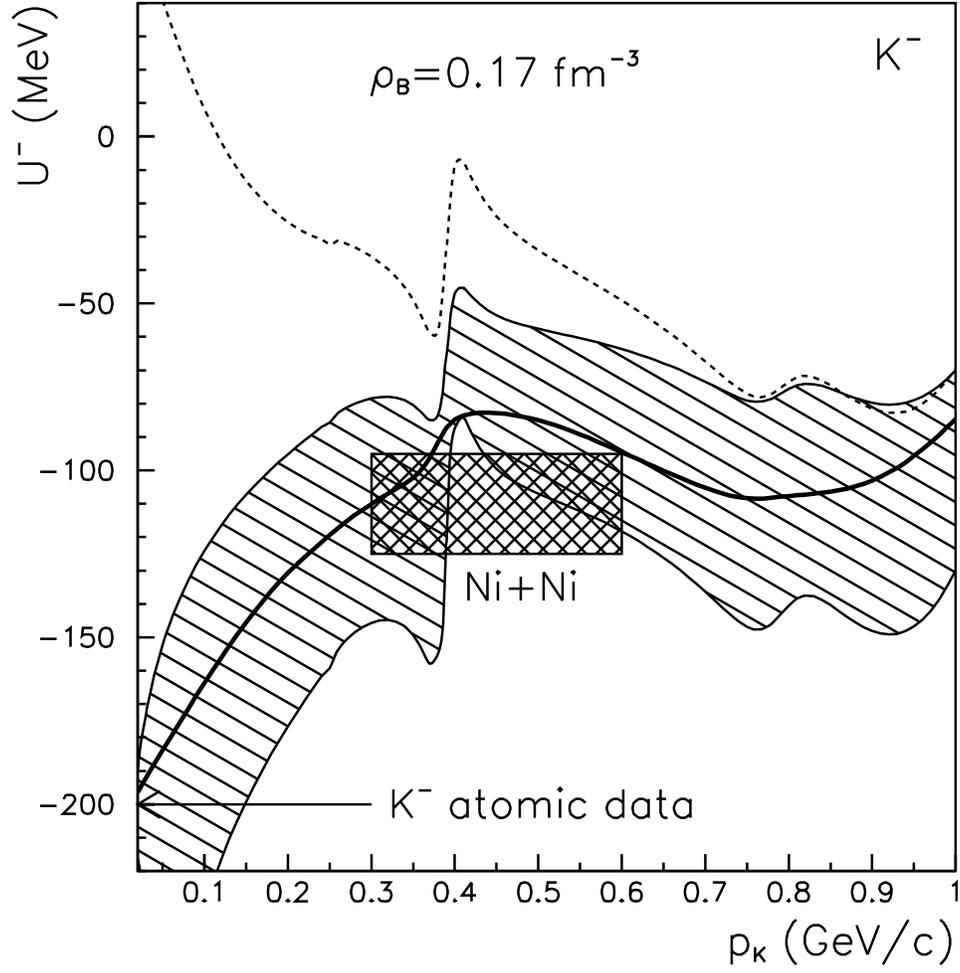,width=15cm}
\caption[]{\label{kmin5}The $K^-$ potential
at nuclear saturation density as a function of the antikaon
momentum $p_K$. The dashed line shows the calculations with the vacuum
$K^-p$ scattering amplitude; the fat solid line is our result
excluding the $\Sigma (1385)$
and $\Lambda (1405)$ resonances while the lined area indicates the
uncertainty of our calculations. The arrow shows the result from the
$K^-$ atomic data~\protect\cite{Friedman} whereas the crossed rectangle
indicates the results from the analysis of $K^-$ production
in $Ni+Ni$ collisions~\cite{Cassing1,Li}.}
\end{figure}

Fig.~\ref{kmin5} illustrates our calculations for the
potential (\ref{forder}) at nuclear saturation density
${\rho}_B={\rho}_0=0.17$~fm$^{-3}$ as a function of the antikaon momentum.
The dashed line indicates the result obtained with the
total free $K^-p$ scattering amplitude while
the fat solid line shows our calculations when excluding the
repulsive low energy resonance contribution in the scattering amplitude. 
The lined area demonstrates the uncertainties due to input parameters. 
In Fig.~\ref{kmin5}  we also show
the antikaon potential as obtained from the analysis of kaonic 
atoms from Friedman, Gal and Batty~\cite{Friedman}, which 
indicates a lowering 
of the effective in-medium ${\bar K}$ mass by $200 \pm 20$~MeV
at zero antikaon momentum. 
The crossed rectangle in  Fig.~\ref{kmin5} shows the
antikaon potential (at $\rho_0$) as implemented in the calculations
in Ref.~\cite{Cassing1}
and Li et al.~\cite{Li97}. The latter potentials have been 
extracted from the experimental data on 
$K^-$ production at ${\theta}_{lab}=0^o$ from $Ni+Ni$ 
collisions at 1.66 and 1.85~A$\cdot$GeV~\cite{Schroter}. 

Note that our result scales the $K^-$ mass
in nuclear matter (at zero momentum) as
\begin{equation}
m_{K^-}^{\ast}=m_{K} \  (1 \ - \ \alpha_K \ {\rho}_B/{\rho}_0) 
\end{equation}
with $\alpha_K \approx 0.2-0.22$ roughly in line with the suggestion by
Brown and Rho~\cite{BrownRho} and with the
recent calculations by Tsushima et al.~\cite{Tsushima}
within the quark-meson coupling model that predicts a decrease of the
$K^-$-meson mass in nuclear matter at ${\rho}_0$ and at zero
momentum by $\simeq 162$~MeV.

\begin{figure}[hbt]
\psfig{file=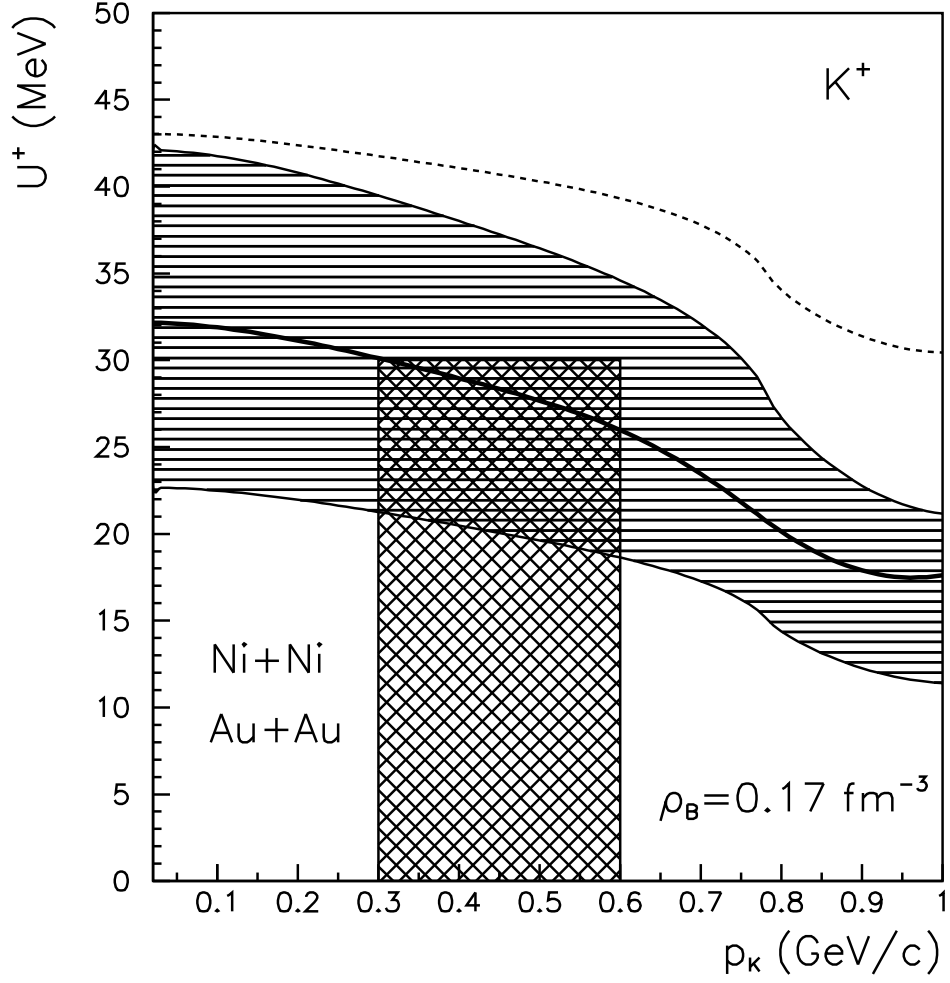,width=15cm}
\caption[]{\label{kmin6}The $K^+$ potential
at nuclear saturation density as a function of the kaon
momentum. The dashed line shows the calculation with the vacuum
$K^+p$ scattering amplitude while the fat solid line is our result
obtained by excluding the $\Sigma (1385)$
and $\Lambda (1405)$ resonances. The lined area demonstrates the
uncertainty of our calculations.  The crossed rectangle
shows the results from the analysis of $K^+$ production
in $Ni+Ni$ and $Au+Au$ collisions at SIS 
energies~\protect\cite{Cassing2}.}
\end{figure}

Fig.~\ref{kmin6} shows the momentum dependence of the 
$K^+$ potential. Again the dashed line indicates our 
calculations using the $K^+p$ forward scattering amplitude in
free space. The solid line is our result obtained when
excluding the low energy solution from Martin~\cite{ADMartin}.
The crossed rectangle in  Fig.~\ref{kmin6} shows the
result of Ref.~\cite{Cassing2}  extracted from the
$K^+$ production data at ${\theta}_{lab}=44^o$ in  $Ni+Ni$ 
collisions at 0.8, 1.0 and 1.8~A$\cdot$GeV~\cite{Barth} and
in $Au+Au$ collisions at 1~A$\cdot$GeV~\cite{Senger}. 
The calculations within the quark-meson coupling 
model~\cite{Tsushima} predict an increase of the
$K^+$-meson mass at ${\rho}_0$ and zero
antikaon momentum by + 24~MeV, which is in reasonable agreement with
our result from the dispersion relation.

\begin{figure}[hbt]
\psfig{file=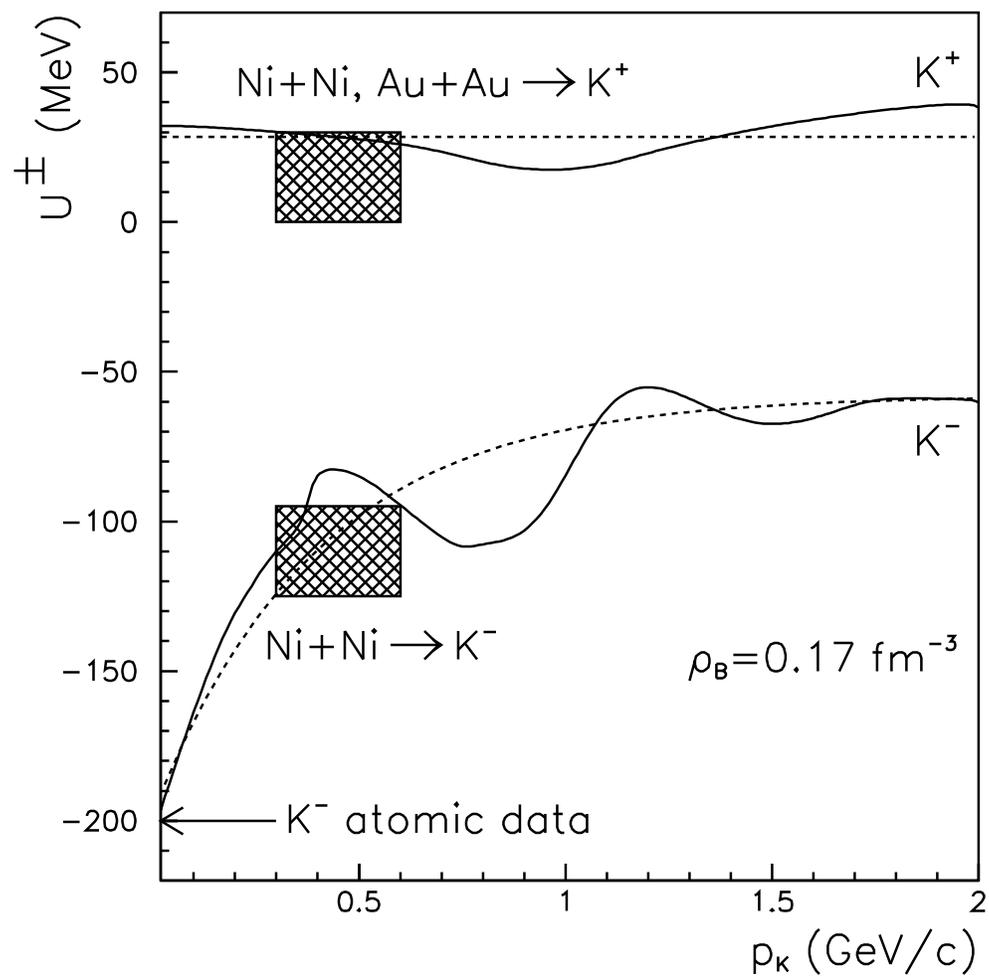,width=15cm}
\caption[]{\label{kmin12}The $K^{\pm}$ potential
at nuclear saturation density as a function of the kaon
momentum. The dashed lines show our parameterization for
the Fermi averaged potentials. The other notations are the same as
in Figs.\protect\ref{kmin5},\protect\ref{kmin6}.}
\end{figure}

Fig.~\ref{kmin12}, furthermore, shows the in-medium correction to the
$K^{\pm}$-meson mass at $\rho_0$ again. All notations are similar to those
in Figs.~\ref{kmin5},\ref{kmin6}.
The dashed lines in Fig.~\ref{kmin12} show our parameterizations
for $U^{\pm}$, which were averaged over the Fermi momentum distribution
at saturation density and read explicitly
\begin{eqnarray}
\label{parap}
U^+({\rho}_B, p_K) &=& - 167 \ {\rho}_B \\
U^-({\rho}_B, p_K) &=& - {\rho}_B \left\lbrace
341+823 \exp (-2.5 p_K) \right\rbrace 
\end{eqnarray}
with the baryon density given in fm$^{-3}$ and $U^{\pm}$ in MeV.

\section{Analysis of data from KEK}
Antikaon production in proton-nucleus collisions is
well suited for the extraction of the
momentum dependence of the $K^-$-meson mass in nuclear matter
because the laboratory $K^{\pm}$ momenta
are relative to the target system which is practically at rest. 
Thus, in principle, by measuring the 
$K^{\pm}$-meson spectrum from $p + A$ collisions one can actually
probe the momentum dependence of the $K^{\pm}$ potential.
According to our studies in Section 3 antikaons of low momenta should see
strong attractive potentials in the nucleus whereas kaons with large 
momenta should only see a slightly attractive potential.

Recently, the $K^-$-meson production in $p+C$ and $p+Cu$
collisions at a beam energy of 3.5 and 4~GeV was studied at 
KEK~\cite{Chiba} where the differential $K^-$ production
cross section was measured at an emission
angle of 5.1$^o$ in the laboratory and at a momentum $p_K=1.5$~GeV/c.
In order to test the sensitivity of the experimental
data to the in-medium $K^{\pm}$ mass we have performed calculations
within the coupled channel transport approach HSD 
used in Refs. \cite{Cassing1,Cassing2} 
for kaon and antikaon production in nucleus-nucleus collisions, in Ref.
\cite{Sib98} for $\rho$-meson production in $p + A$ collisions and
in Ref. \cite{pbar} for antiproton production in $p + A$ and 
$A + A$ collisions
at SIS energies. We note that the local density approximation
involved in the transport calculation as well as the on-shell treatment
of production and absorption events might be questionable 
for a light nucleus
such as $^{12}C$, but the detailed comparison to experimental 
pion spectra performed in Ref. \cite{Sib98} and to $K^+$ spectra in
Ref. \cite{Varenna} indicates that these uncertainties should be less than
30\%. This systematic error (arising from the transport simulation itself)
is smaller than the uncertainty in the elementary $K^-$ production cross
section in $pN$ reactions (cf. Section 2). 

\begin{figure}[h]
\psfig{file=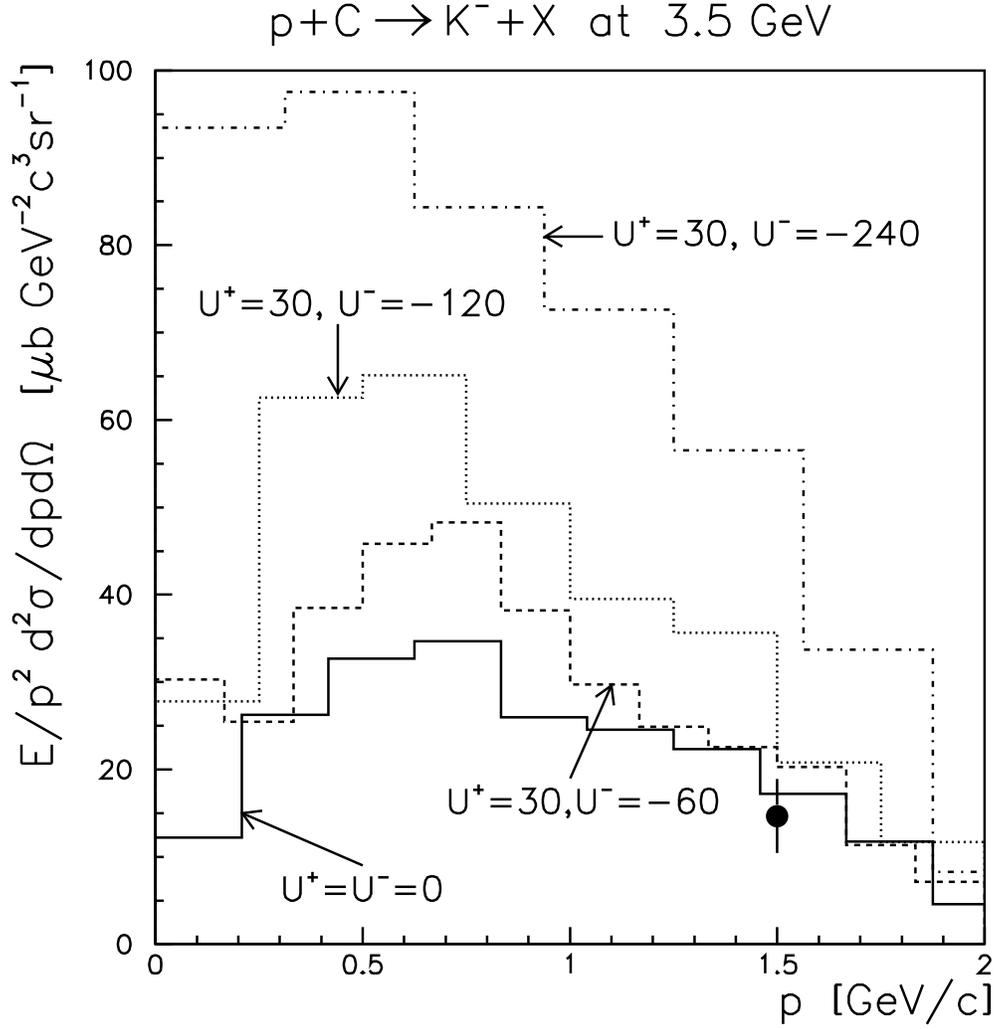,width=15cm}
\caption[]{\label{kmin13}Momentum spectra of 
$K^-$-mesons from $p + ^{12}C$ collisions at a bombarding energy of 
3.5~GeV and laboratory angle $\theta_{lab}=5^o$
calculated for different in-medium potentials
for kaons $U^+$ and antikaons $U^-$ in MeV. The experimental point 
is taken from Ref.~\protect\cite{Chiba} and shown with
the systematical error.}
\end{figure}

Fig.\ref{kmin13} shows the $K^-$ spectra from $p + ^{12}C$ collisions
at 3.5~GeV calculated for different values of $U^{\pm}$
in comparison to the experimental data point from Ref. \cite{Chiba}
shown with both the statistical and systematical errors. 
Actually, at an antikaon momentum of 1.5 GeV/c one can not 
distinguish very well between the calculations with bare masses and e.g.
$U^+$=30~MeV and $U^-=-$60~MeV as obtained in Section 3.
Howewer, it seems clear that the experimental point allows to exclude a
very strong attractive $K^-$ potential of e.g. $U^-= -$240~MeV. 
On the other hand our calculations indicate a substantial change of the
$K^-$-meson spectra due to in-medium modifications especially at low
antikaon momenta.

\begin{figure}[h]
\psfig{file=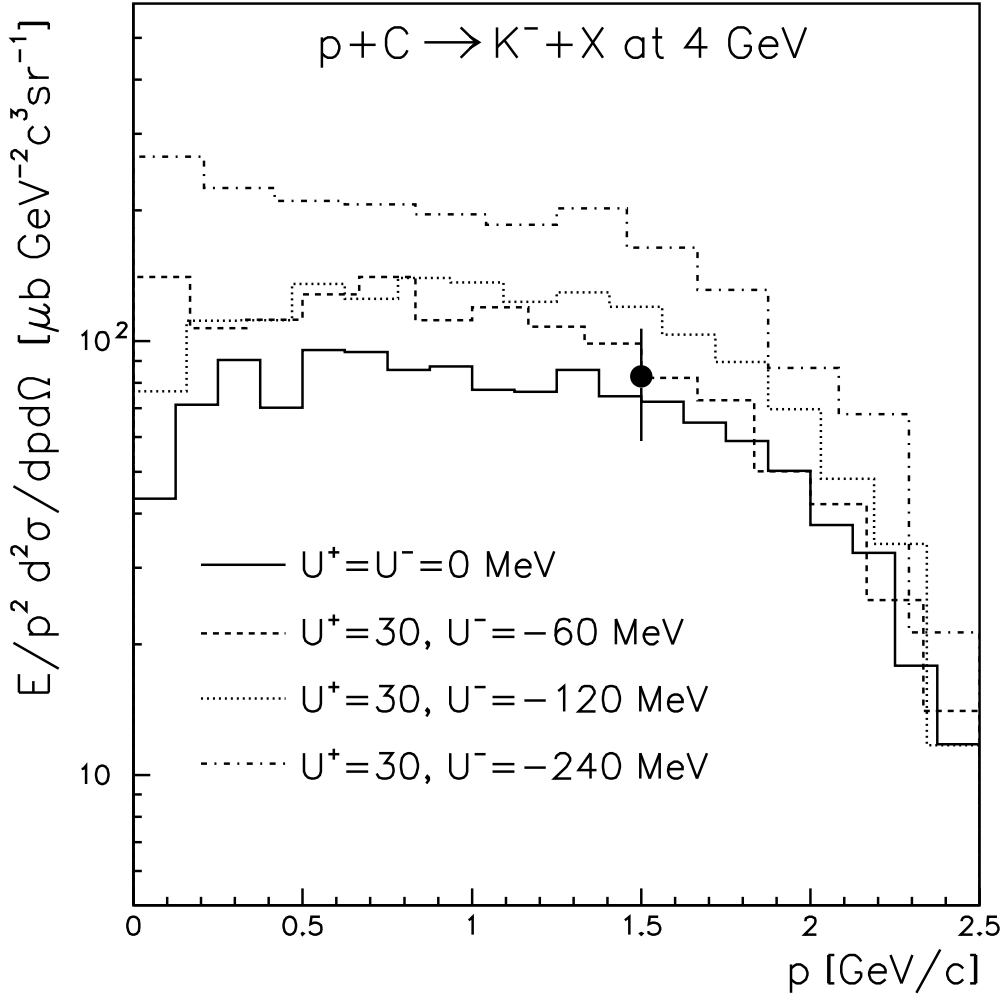,width=15cm}
\caption[]{\label{kmin10}The momentum spectra of 
$K^-$-mesons from $p+C$ collisions at the bombarding energy of 
4~GeV. The histograms show our
calculations with different in-medium potentials
for kaons $U^+$ and antikaons $U^-$ in MeV. The experimental point 
is taken from Ref.~\protect\cite{Chiba} and shown with
the systematical error.}
\end{figure}

\begin{figure}[h]
\psfig{file=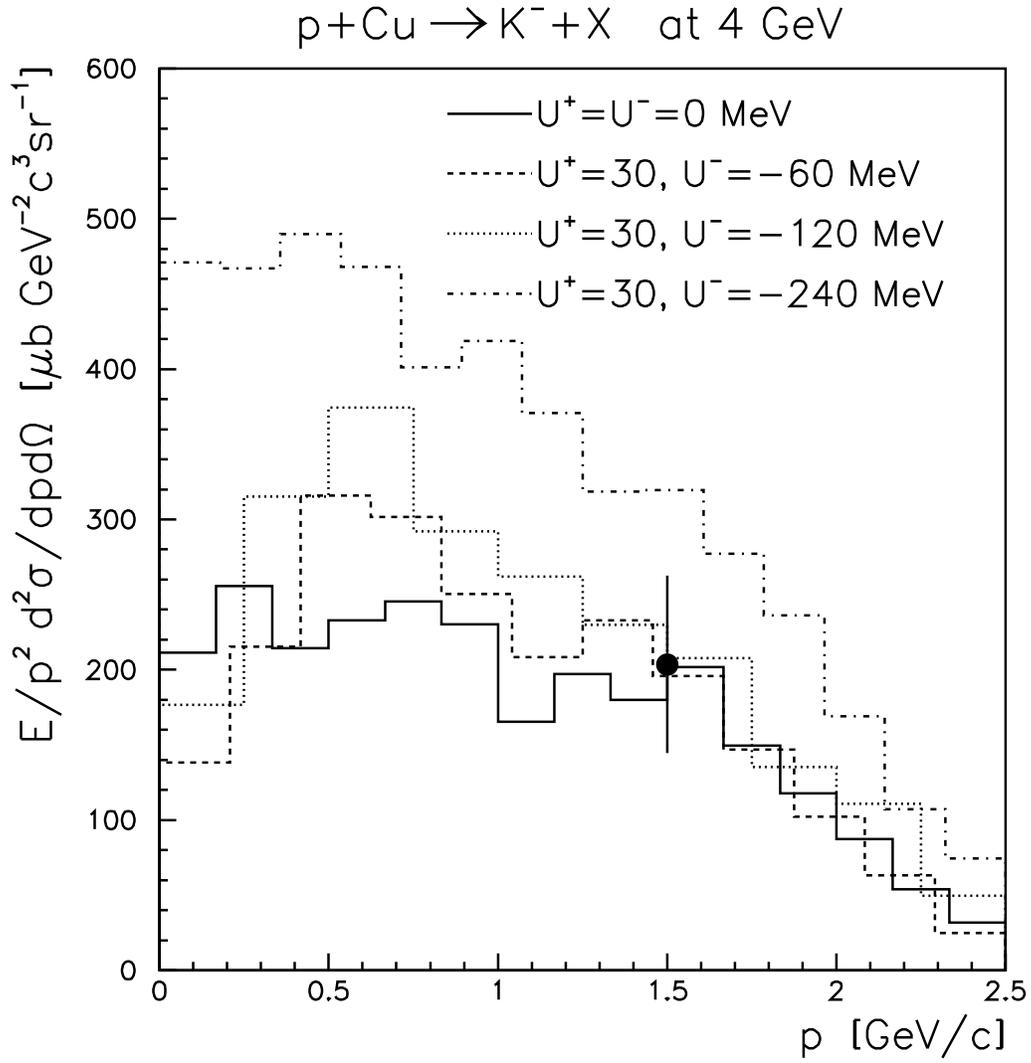,width=15cm}
\caption[]{\label{kmin9}The momentum spectra of 
$K^-$-mesons from $p+Cu$ collisions at the bombarding energy of 
4~GeV. The histograms show our
calculations with different in-medium potentials
for kaons $U^+$ and antikaons $U^-$ while the full dot indicates the 
experimental data point from Ref.~\protect\cite{Chiba}.}
\end{figure}

Figs. \ref{kmin10},\ref{kmin9} show our calculations
for the momentum spectra of $K^-$-mesons from
$p+C$ and $p+Cu$ collisions at the bombarding energy of 4~GeV
for different values of the kaon and antikaon potential $U^\pm$ at
$\rho_0$ in MeV.
Though the sensitivity to kaon and antikaon potentials is less pronounced
at the higher energy for both systems we may conclude 
that the experimental data do not
contradict an in-medium modification for $K^+$ and $K^-$
mesons according to the momentum-dependent $K^{\pm}$ potential.
Furthermore, the fast antikaons are not sensitive to the actual value
of the in-medium potential $U^{\pm}$, since we
do not observe sizeable differences between the
calculations with bare $K^+$ and $K^-$ masses and with
potentials   $U^+$=+30~MeV and $U^-=-$60~MeV as expected from Section 3. 

On the other hand, our analysis of $K^-$-production with
laboratory momenta of 1.5~GeV excludes a strong
reduction of the antikaon mass for $U^-\le 120~MeV$,
which was found for low energy antikaons (in the cms) from
heavy-ion collisions. Moreover, 
Figs.~\ref{kmin13},\ref{kmin10},\ref{kmin9}
illustrate, that in case of proton-nucleus collisions the
low energy antikaons are most sensitive to in-medium effects. This
effect can also be understood easily when comparing the ratio of the
kaon or antikaon potential with the kaon energy at a fixed momentum. This 
ratio (even for momentum-independent potentials) increases with 
decreasing momentum. Simply due to energy conservation an antikaon seeing
an attractive potential in the nucleus will be decelerated on its way
to the continuum thus enhancing the spectrum at low momenta - provided
that it is not absorbed in the nucleus via the reaction $K^- N
\rightarrow \pi Y$, where $Y$ denotes a hyperon.

\section{Predictions for COSY energies}
The in-medium change of the $K^-$ properties
can be observed through an enhancement
of the production yield or, as found before,   
in the modification of the antikaon spectrum.
Both phenomena might be studied in proton-nucleus reactions
experimentally and constrain the information about the
$K^{\pm}$ potential in nuclear matter at normal 
density ${\rho}_0$. Whereas the data points at KEK are above the threshold
in free nucleon-nucleon collisions, one expects that in-medium potentials
are more pronounced at subthreshold energies in $p + A$ reactions
as in case of antiproton production \cite{pbar}.

Experiments on $K^-$ production in $p+A$ collisions have
been proposed for the proton synchrotron COSY~\cite{Muller}.
The $K^-$-meson production cross section as well as the
momentum spectra will be measured for 
bombarding energies around 2.5~GeV and kaon emission angles
$\theta_{lab} \le 10^o$. 
In these experiments a rather precise measurement of
the $K^-$-meson spectra can be performed, which might allow to reconstruct
the antikaon potential up to momenta of 1.3~GeV/c~\cite{SibirtsevR}.
The $K^-$-meson production threshold in $NN$ collisions in 
free space corresponds to a proton beam energy of
$\simeq 2.5$~GeV and the experiments at COSY aim at
the near or subthreshold physics. Actually the cross section
for subthreshold $K^-$ production in proton-nucleus collisions 
is small but the reaction threshold might drop due to the
decrease of the in-medium antikaon mass \cite{Varenna}.

\begin{figure}[h]
\psfig{file=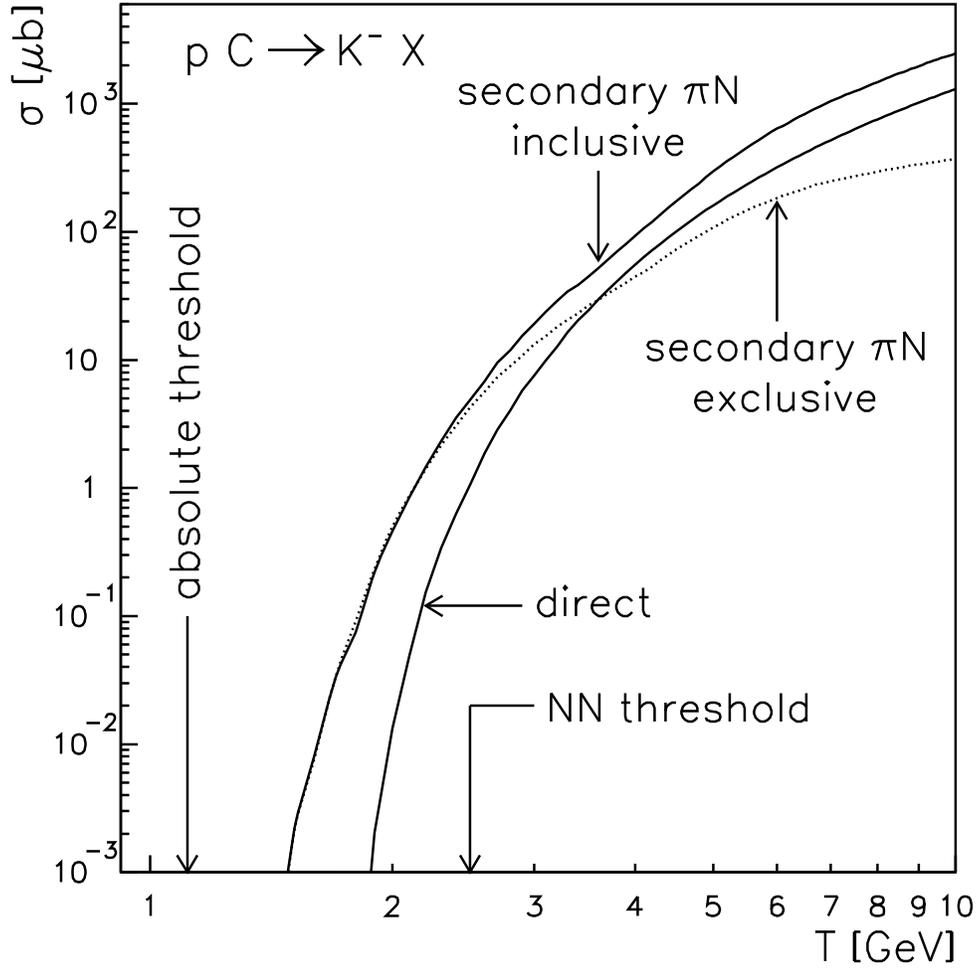,width=15cm}
\caption[]{\label{kmin11}The cross section for $K^-$ production
in $p+^{12}C$ collisions as a function of the bombarding energy
using bare kaon and antikaon masses.
The lines show our calculations for the direct ($pN$) and secondary
($\pi N$) production mechanisms. The arrows indicate the 
absolute and $NN$  reaction threshold, respectively.}
\end{figure}

Fig.~\ref{kmin11} shows the  $K^-$ production cross section integrated 
over all antikaon momenta from $p+^{12}C$ collisions as
a function of the bombarding energy. The lines indicate our 
calculations with bare $K^{\pm}$ masses and
for the separate contributions from the direct $pN$ and
secondary $\pi N$ production channel. For the
$K^-$-meson production from secondary $\pi N$ interactions
we show the calculations using only the exclusive 
$\pi N \to N K K^-$ channel (dotted line) as well as the 
inclusive $\pi N \to K^- X$ reaction channels as 
illustrated in Fig.~\ref{kmin7}.
Note, that while at KEK energies the contributions from
the direct and secondary  $K^-$ production mechanism are 
approximately the same, at COSY energies the secondary reaction channels
are dominant. In Fig.~\ref{kmin11} we also indicate the proton beam
energy corresponding to the absolute threshold for the antikaon 
production in $p+C$ collisions, which is available for the
coherent process $pC \to p C K K^-$.

\begin{figure}[h]
\psfig{file=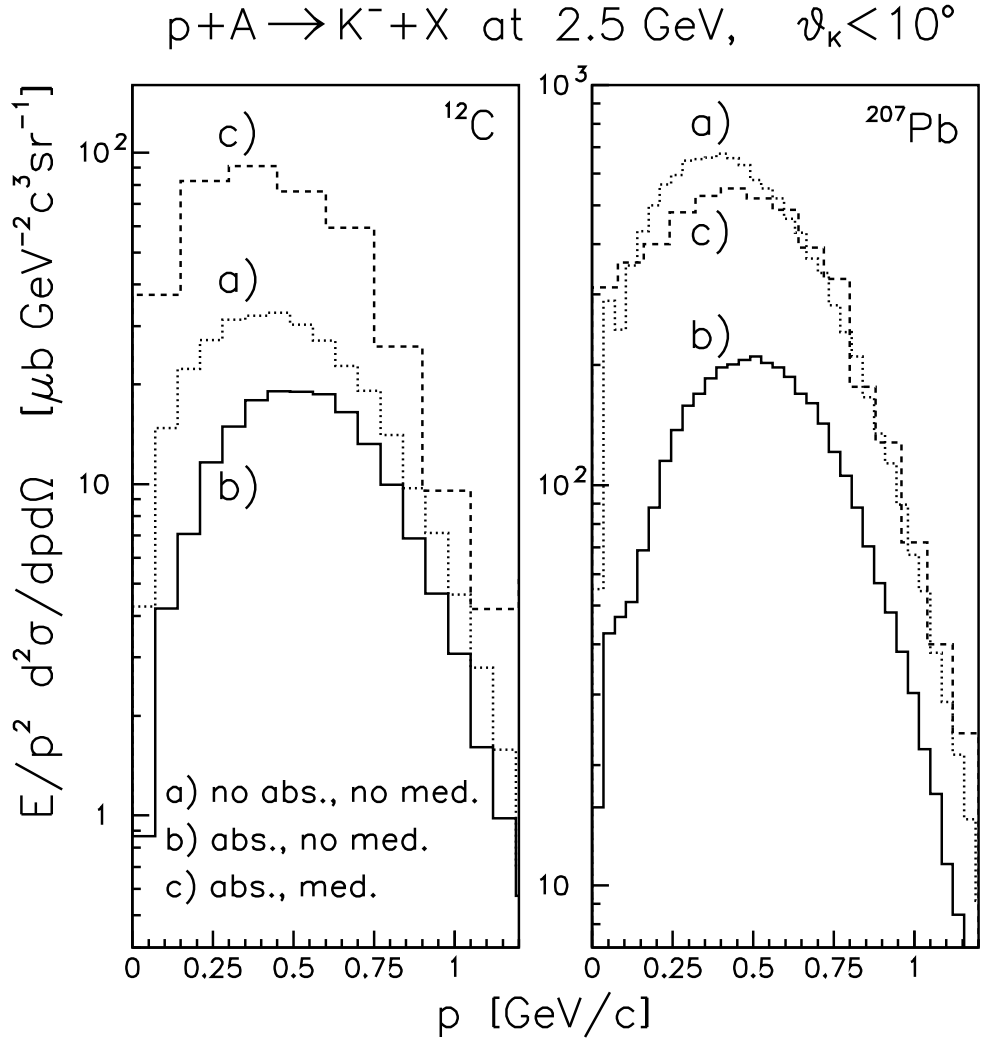,width=15cm}
\caption[]{\label{kmin16}The $K^-$ spectrum from
$p+^{12}C$ and $p+^{207}Pb$ collisions at 2.5~GeV for antikaon emission
angles $\leq$ 10$^o$. The solid histograms show our calculations
with bare masses and $K^-$ absorption (b), the dotted
histograms  are calculated with bare masses and without
absorption (a) while the dashed histogram is our result for  
in-medium potentials of $K^+$ ($U^+=30$ MeV) and $K^-$ 
($U^-=-120$ MeV) mesons and with $K^-$ absorption (c).}
\end{figure}

Fig.~\ref{kmin16}, furthermore, shows the 
calculated momentum spectrum for $K^-$-mesons
produced in $p+^{12}C$ and $p + ^{207}Pb$ 
collisions at the proton beam energy of 2.5~GeV.
The spectrum is integrated over the $K^-$ emission angle up
to $\theta_{lab} =10^o$, which corresponds to the experimental
setup of the COSY-ANKE detector. The solid histograms (b) show
our calculation with bare masses for both kaon and
antikaon. The dotted histograms (a) show the calculations with
bare masses when neglecting the final state absorption 
of $K^-$-mesons. It is clear that FSI have a very strong effect
on the antikaon spectrum. $K^-$ absorption gives roughly a factor of
2 reduction in cross section for $^{12}C$ while it amounts to a 
factor $\approx$ 5 for $^{207}Pb$. 
The dashed histograms (c) indicate our results 
obtained with in-medium potentials
$U^+=30$ and $U^-=-120$~MeV and taking into account
the FSI.  In the latter case the absolute $K^-$-meson production
cross section substantially increases and the spectrum is shifted 
to lower momenta as noted before.

We conclude that the shape of the $K^-$ spectrum as well as 
the absolute value of the production cross section from
proton-nucleus collisions are very
sensitive to the magnitude of the $K^{\pm}$ potential in the
baryonic matter and should be determined experimentally by comparing
the spectra from light and heavy targets as a function of momentum.

\section{Conclusions}
We have calculated the momentum dependence of the $K^+$ and
$K^-$-meson potentials based on dispersion relations for the
$K^{\pm}N$ forward scattering amplitude at finite nuclear density.
We have found that the in-medium  $K^+$ potential almost does
not depend on the kaon momentum and increases to $\simeq +30$~MeV
at nuclear matter density ${\rho}_B=0.17$~fm$^{-3}$.
The $K^-$ potential drops to $\simeq -200$~MeV at zero antikaon momentum
and saturates at $ \approx -60$~MeV at high momenta.
Our results are in reasonable agreement with the 
in-medium potentials evaluated from the measurement of kaonic 
atoms~\cite{Friedman}, as well as from the previous 
studies~\cite{Li,Cassing1,Cassing2} on $K^{\pm}$-production from 
heavy-ion collisions~\cite{Schroter,Barth,Senger}.

Furthermore, we have analysed the recent data from KEK~\cite{Chiba} 
on antikaon production in
proton-nucleus collisions at  bombarding energies of 3.5 and
4~GeV. The data do not contradict our results for the 
momentum dependent $K^{\pm}$ potential, however, the experimental 
results are not very sensitive to the actual magnitude of the
in-medium masses at the high momentum of 1.5 GeV.
On the other hand, we have illustrated that $p + A$ reactions
at COSY for bombarding energies close to the threshold in proton-proton
collisions might be well suited for the evaluation of the
$K^{\pm}$ potential in nuclear matter (at $\rho_0$) as a function of
the antikaon momentum in the laboratory.

\section*{Appendix}
The total cross section for the $K^+ p$ reaction can be parametrized
as a function of the kaon momentum $p_K$ in the laboratory
system as
 
\begin{figure}[h]
\psfig{file=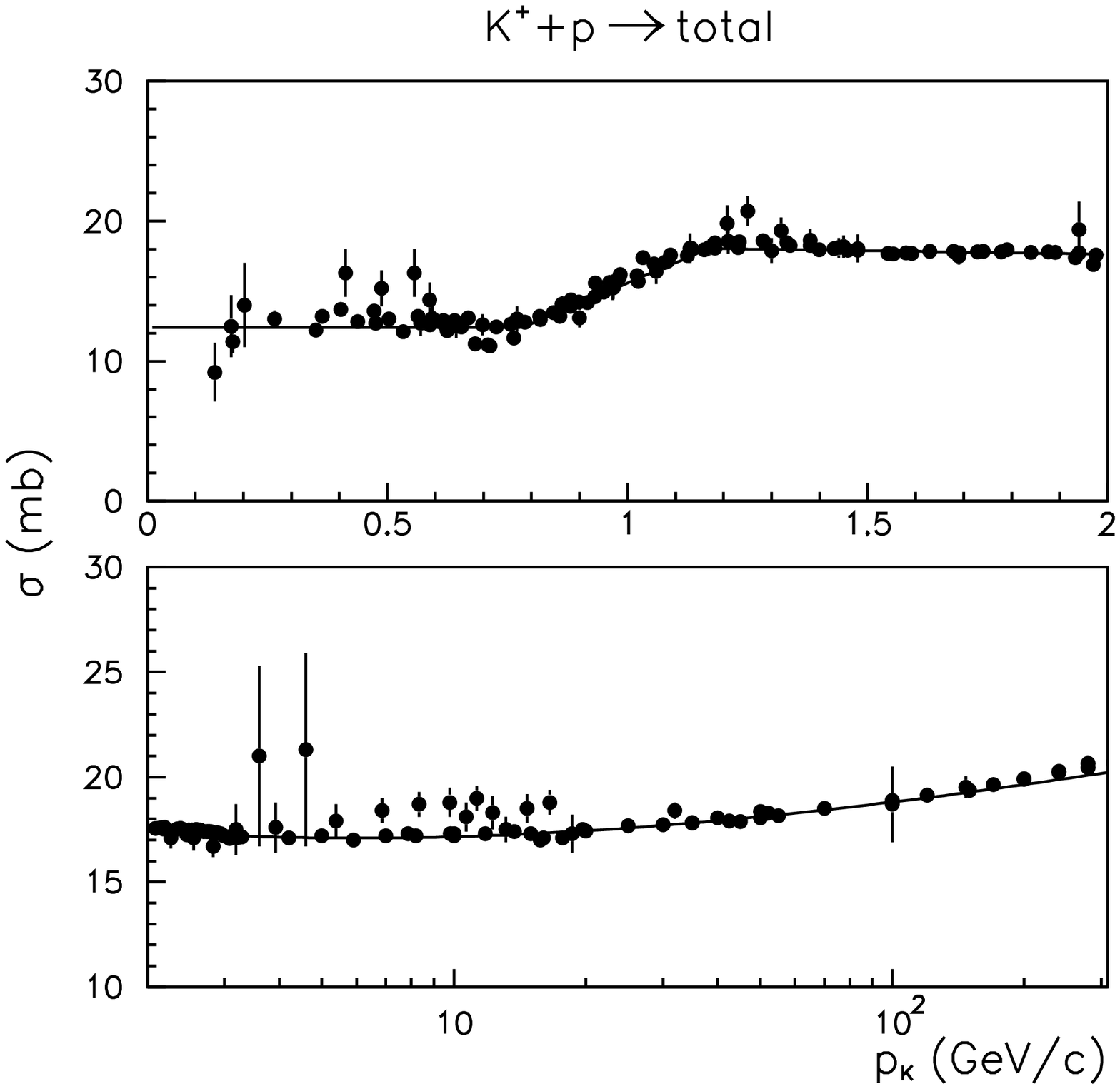,width=15cm}
\caption[]{\label{kmin1}Total $K^+p$ cross section as a
function of the kaon momentum in the laboratory frame.
The experimental data are taken from Ref.~\protect\cite{LB} while the
lines show our parametrization.}
\end{figure}

\vspace*{0.17cm}
$\begin{array}{llr}
\sigma = 12.4            & 
\hspace*{2cm}  p_K \le 0.78 & \\ 
\sigma = 1.09+ 14.5 \  p_K   & 
\hspace*{2cm}  0.78 < p_K \le 1.17 &
\hspace*{4.8cm} (23) \\
\sigma = 18.64 - 0.5 \ p_K  & \hspace*{2cm}  1.17 < p_K \le 2.92 &  
\end{array}$ 

\noindent 
where the cross section is given in $mb$ and the
kaon momentum $p_K$ in $GeV/c$.
At higher momenta we use the Regge parametrization from 
Donnachie and Landshoff~\cite{Donnachie}. Fig.~\ref{kmin1}
illustrates the comparison between our parametrization
and the experimental data~\cite{LB}.

The total $K^- p$ cross section is parametrized as

\vspace*{0.17cm}
$\begin{array}{llr}
\sigma = 26.32/p_K &  
\hspace*{2cm} p_K \le 0.25 & \\
\sigma = 36.4 \ p_K^{-0.26} & 
\hspace*{2cm} 1.81 < p_K \le 3  & \hspace*{5.9cm} (24) \\
\sigma = 32.25 \  p_K^{-0.15} & 
\hspace*{2cm} 3 < p_K \le 17  & \\ 
\end{array}$ 

\begin{figure}[h]
\psfig{file=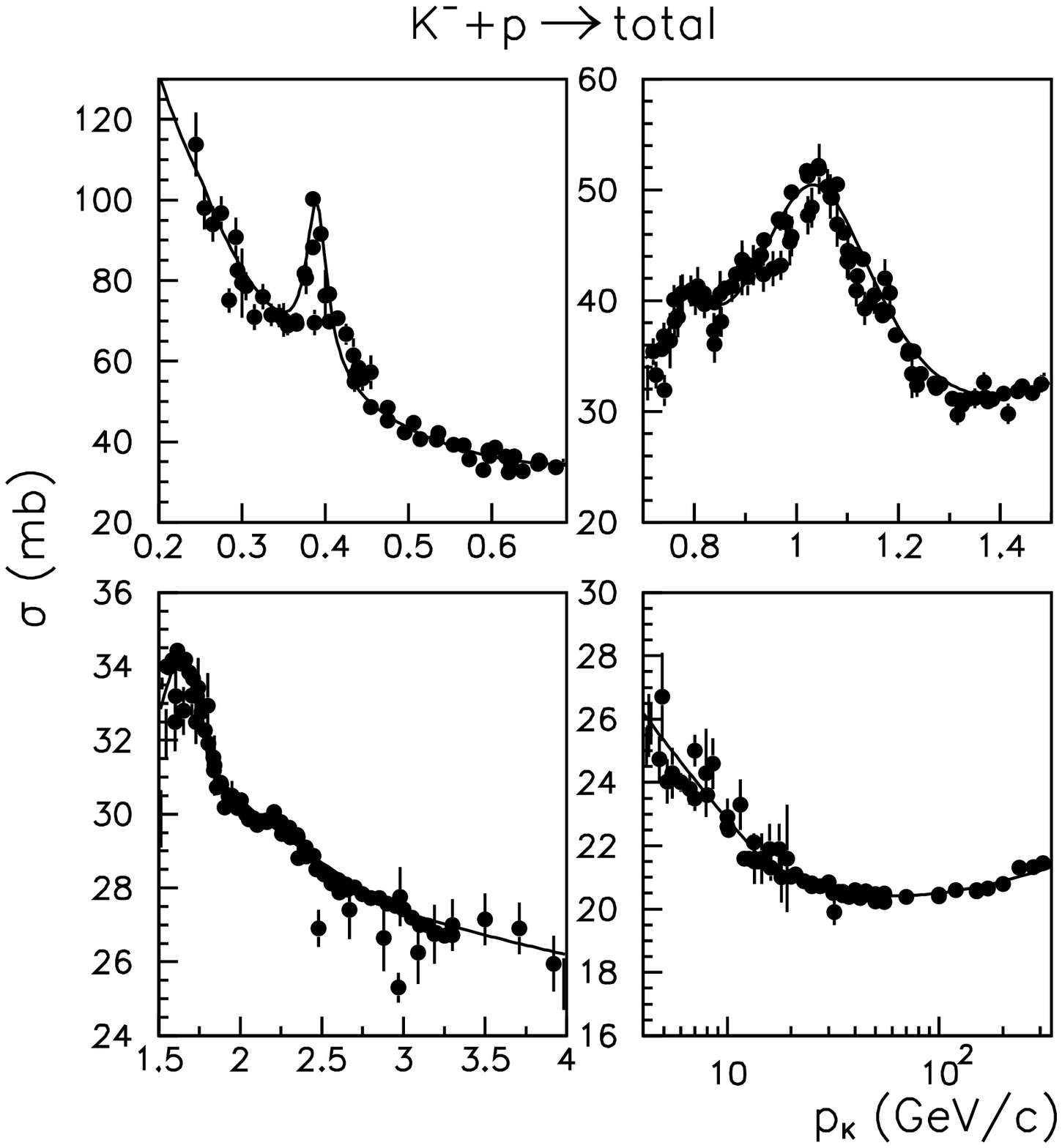,width=15cm}
\caption[]{\label{kmin2}Total $K^-p$ cross section as
a function of the kaon momentum in the laboratory frame.
The experimental data are taken from Ref.~\protect\cite{LB} 
while the lines show our parametrization.}
\end{figure}

In the antikaon momentum range from 0.25 to 1.81~GeV/c
the cross section is approximated as
\[
\hspace*{0.9cm} \sigma = 13.2 p_K^{-1.47} + \sum_{i=1}^4
C_i \ \frac{ {\Gamma}_i^2}{(\sqrt{s}- M_i)^2+{\Gamma}_i^2/4}
\hspace*{6.5cm} (25)
\]
where the parameters $C_i$, ${\Gamma}_i$ and $M_i$ are listed in 
Table~\ref{tabp} and $\sqrt{s}$ denotes the invariant
mass of the colliding $KN$ system. Again at high $K^-$ momenta
we use the results from Ref.~\cite{Donnachie}.
Fig.~\ref{kmin2} shows our parametrization together with the
experimental data from Ref.\cite{LB}.

\begin{table*}[h]
\begin{center}
\caption{\label{tabp}Parameters of the 
approximation~(25).}
\vspace{0.6cm}
\begin{tabular}{|l|c|c|c|}
\hline
$i$ & $C_i$ (mb) & $M_i$ (GeV) & ${\Gamma}_i $ (GeV)\\ 
\hline
1 & 10.4 & 1.518 & 0.012 \\
2 & 7.6 & 1.81 & 0.17 \\
3 & 1.8 & 1.695 & 0.05 \\
4 & 6.4 & 2.107 & 0.37 \\
\hline 
\end{tabular}
\end{center}
\end{table*}

\newpage

\begin{figure}[hbt]
\psfig{file=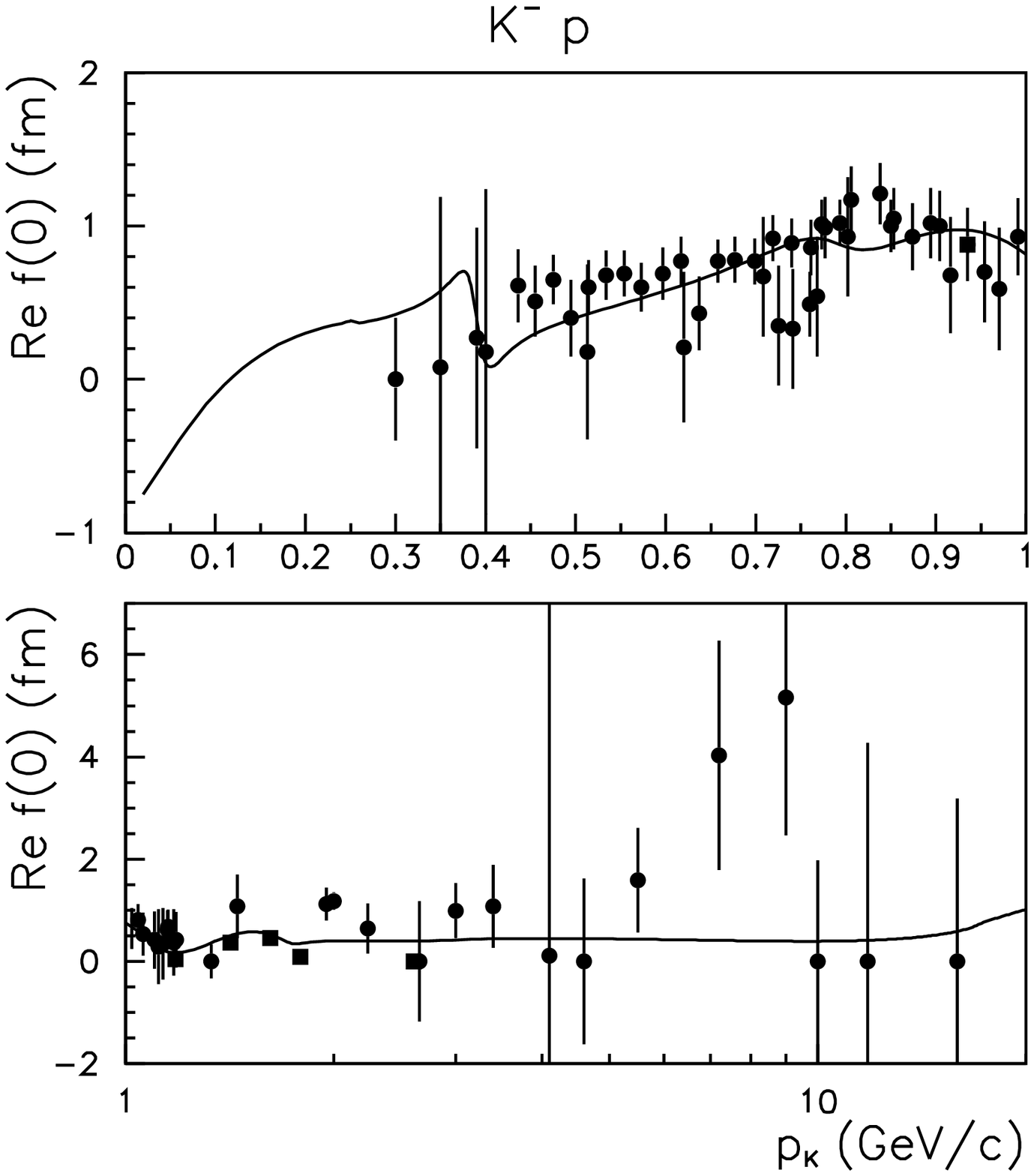,width=15cm}
\caption[]{\label{kmin3}Real part of the $K^-p$
scattering amplitude (solid line) in comparison to the 
experimental data taken from Ref.~\protect\cite{Dumbrais}.}
\end{figure}

\end{document}